\documentclass[12pt,a4paper]{article}
\usepackage{a4wide}
\usepackage{latexsym}
\usepackage{epsf}
\usepackage{amssymb}
\begin{document}
\def\be{\begin{equation}}
\def\ee{\end{equation}}
\def\bea{\begin{eqnarray}}
\def\eea{\end{eqnarray}}

\def\pd{\partial}
\def\a{\alpha}
\def\b{\beta}
\def\bi{\begin{itemize}}
\def\ei{\end{itemize}}
\def\g{\gamma}
\def\d{\delta}
\def\m{\mu}
\def\n{\nu}
\def \h{\mathcal{H}}
\def \hh{\mathcal{G}}
\def\t{\tau}
\def\p{\pi}
\def\th{\theta}
\def\l{\lambda}
\def\cd{\cos{y_2}}
\def\cu{\cos{y_1}}
\def\su{\sin{y_1}}
\def\sd{\cos{y_2}}

\def\O{\Omega}
\def\r{\rho}
\def\s{\sigma}
\def\e{\epsilon}
  \def\scri{\mathcal{J}}
\def\cM{\mathcal{M}}
\def\tcM{\tilde{\mathcal{M}}}
\def\RR{\mathbb{R}}

\hyphenation{re-pa-ra-me-tri-za-tion}
\hyphenation{trans-for-ma-tions}


\begin{flushright}
IFT-UAM/CSIC-04-02\\
gr-qc/0401097\\
\end{flushright}

\vspace{1cm}

\begin{center}

{\bf\Large   The infinite curvature limit of AdS/CFT.}\footnote{
Contribution to the proceedings of the 2003 ``Peyresq Physics 8'' meeting.}

\vspace{.5cm}

{\bf Enrique \'Alvarez }

\vspace{.3cm}

\vskip 0.4cm  
 
{\it  Instituto de F\'{\i}sica Te\'orica UAM/CSIC, C-XVI,
and  Departamento de F\'{\i}sica Te\'orica, C-XI,\\
  Universidad Aut\'onoma de Madrid 
  E-28049-Madrid, Spain }

\vskip 0.2cm

\vskip 1cm

{\bf Abstract}

\end{center}

\begin{quote}
 Some kinematical speculations on the infinite curvature limit of the conjectured duality of Maldacena
  between ten-dimensional strings living in $AdS_5\times S_5$ and a ordinary four-dimensional quantum field theory, namely 
${\cal{N}}=4$ super Yang-Mills with gauge group
$SU(N)$ are given.
\end{quote}


\newpage

\setcounter{page}{1}
\setcounter{footnote}{1}
\newpage
\section{Introduction}
The usual AdS/CFT correspondence relates strings living on a manifold of curvature
\be
R\sim\frac{1}{l^2}
\ee
to a ordinary four-dimensional conformal field theory (CFT) with gauge coupling $g$ given in terms
of the string coupling constant, $g_s$, by
\be
g=g_s^{1/2}
\ee
The  't Hooft coupling is 
\be
\l\equiv g^2 N\equiv g_s N
\ee
Both the 't Hooft coupling and the effective string tension are
 given in terms of the string length $l_s^2\equiv \a^{\prime}$ by
\be
\l^{1/2}=\frac{l^2}{l_s^2}\sim T_{eff}
\ee
and the ten-dimensional Newton constant is
\be
\kappa_{10}^2\sim l_p^8=g_s^2 l_s^8\sim \frac{l^8}{N^2}
\ee
The correspondence is usually studied in the low curvature regime in which
\be
\frac{l}{l_s}>>1
\ee
In this regime strings are believed to be well approximated by supergravity, but the CFT is strongly coupled.
Many nontrivial checks are however possible owing to the existence of gauge invariant operators protected
by supersymmetry whose correlators are total or partially determined through kinematics. On the string side
these correlators are determined by computing the action for the relevant fields in terms of arbitrary sources
at the conformal boundary.
\par
The opossite limit, i.e.
\be
\frac{l}{l_s}<<1
\ee
corresponds to effectively tensionless strings in a strongly curved (and, as we shall see, somewhat singular)
background. The corresponding CFT is, however, in the perturbative regime. 
\par
On  the string theory side, however,
it is not even clear what is the meaning of the sources, and it is not known how to decode the information suposedly 
provided by the preturbative CFT. 
\par
The purpose of the present note is the quite modest one of discussing the high curvature limit of $AdS$, and to argue
that it is none other than the light cone itself.
\section{The infinite curvature limit of constant curvature spaces}
Constant curvature spaces of any signature can be understood (cf., for example, \cite{Alvarez})
as hypersurfaces of flat n-dimensional space with metric
\be
ds^2 = \sum_{a=1}^n\e_a dx_a^2
\ee
where all $\e_a=\pm 1$. The signs are arbitrary, {\em except} for the condition that at least one coordinate, but not all of them, 
 has got to be
timelike, which in our conventions means positive sign. 
\par
Calling $x_{n-1}$ one of the timelike coordinates, and $x_n$ one of the spacelike ones, this means that the metric enjoys the term
\be
dx_{n-1}^2-dx_n^2
\ee
The equation deternmining the surface itself is
\be
\sum_{a=1}^n \e_a x_a^2 = \pm l^2
\ee
Here the length scale $l$ determined the curvature through
\be
R=\pm \frac{n(n+1)}{l^2}
\ee
All these manifolds enjoy a maximal group of isometries, which is a real form of $SO(n)$. The Killings are given by
\be
L_{ab}\equiv \e_a x^a \pd_b-\e_b x^b\pd_a
\ee
(no Einstein implicit sum convention is applied in this definition).
Horospheric coordinates are defined by
\bea
&&x_{-}\equiv x_n-x_{n-1}\nonumber\\
&&z\equiv \frac{l}{x_{-}}\nonumber\\
&&y^i\equiv z x^i\,i=1\ldots n-2.
\eea
The metric reads in general
\be
ds^2=\frac{\sum_i \e_i dy_i^2 \mp l^2 dz^2}{z^2}
\ee
The case corresponding to our present interest is when 
\be
\e_i=-1\,\forall i
\ee
It has isometry group $SO(1,n)$, and metric
\be
ds^2=\frac{-\sum_i  dy_i^2 \mp l^2 dz^2}{z^2}
\ee
The lorentzian form is de Sitter space, and the euclidean form is what is usually called
{\em euclidean Anti de Sitter}; although it could equally well be called {\em euclidean de sitter}.
(There are no euclidean versions with isometry group $SO(2,n)$).

\par
Written in this form, it is quite obvious that when $l\rightarrow 0$ (which corresponds to the equation
\be
x_{n-1}^2-\sum_{i=1}^{n-2}x_i^2-x_n^2 =0
\ee
is the light cone of the origin in ordinary n-dimensional Minkowski space, which we will denote
by $N_{\pm}(0)$, with metric
\be\label{degenerada}
ds^2=\frac{-\sum_i  dy_i^2}{z^2}
\ee

\section{Life on the light cone}
The local structure of the light cone is $S^{n-2}\times\mathbb{R}^{+}$, and a point in $N_{+}$
can be specified by $(x_0,n^i)$, where $x_0\in \mathbb{R}^{+}$ and $\vec{n}^2=1$ is a point
on the unit $(n-2)$-dimensional sphere, $S_{n-2}$, that is, a $(n-1)$-dimensional structure. 
The light cone can be visualized as a $S_{n-2}$ sphere of radius $x_0$.
\par

The induced metric $h_{ij}$ is, however, 
degenerate (that is, as a matrix it has rank $n$), because
the time differential is totally absent from the line element:
\be
ds_{+}^2=x_0^2 d\Omega_{n-2}^2
\ee
where $d\Omega_n^2$ is the metric on the unit n-sphere, $S_n$, which in terms of
angular variables reads: 
\be\label{cono}
d\Omega_n^2\equiv d\theta_n^2 + \sin{\theta_n}^2 d\theta_{n-1}^2+\ldots + \sin{\theta_n}^2
\sin{\theta_{n-1}}^2\ldots \sin{\theta_2}^2 d\theta_1^2
\ee
This means that, although singular as a metric on $N_{+}$, the metric is perfectly regular 
(actually
the standard one) as a metric on the (n-2)-spheres $t=constant$.
\par

The invariant volume element, however, vanishes, due to the fact that
\be
\sqrt{h}=0
\ee

Remarkably enough, and in spite of some statements on the contrary, the complete set of isometries of the 
{\em three}-dimensional metric (\ref{cono}) includes the full four-dimensional Lorentz group,
$SO(1,3)$. \footnote{Isometries are well-defined, even for singular metrics, through the vanishing Lie-derivative condition
$\pounds (k) g_{\m\n}=0$, reflecting the invariance of the metric under the corresponding one-parametric group 
of diffeomorphisms, although of course this is not equivalent to $\nabla_{\m} k_{\n}+\nabla_{\n} k_{\m}=0$ because the 
covariant derivative (that is, the Christoffel symbols ) is not well defined owing to the absence of the inverse metric.
}This should be quite obvious from our limiting process of the previous paragraph.
\par
The six Killing vectors with factorizable coeficients which generate $SO(1,3)$ are actually given  by: 
\bea
&&J_1=\cos{\phi}\frac{\pd}{\pd\theta}-\cot{\theta}\sin{\phi}\frac{\pd}{\pd \phi}\nonumber\\
&&J_2=\sin{\phi}\frac{\pd}{\pd\theta}+\cot{\theta}\cos{\phi}\frac{\pd}{\pd \phi}\nonumber\\
&&J_3=\frac{\pd}{\pd \phi}\nonumber\\
&&K_1= - 0 \sin{\theta}\sin{\phi}\frac{\pd}{\pd x^0}-\cos{\theta}\sin{\phi}\frac{\pd}{\pd \theta}
-\frac{\cos{\phi}}{\sin{\theta}}\frac{\pd}{\pd \phi}\nonumber\\
&&K_2=x^0 \sin{\theta}\cos{\phi}\frac{\pd}{\pd x^0}+\cos{\theta}\cos{\phi}\frac{\pd}{\pd \theta}
-\frac{\sin{\phi}}{\sin{\theta}}\frac{\pd}{\pd \phi}\nonumber\\
&&K_3=x^0 \cos{\theta}\frac{\pd}{\pd x^0}-\sin{\theta}\frac{\pd}{\pd \theta}.
\eea
What it is perhaps not immediatly obvious is that this is not the full history; it will be shown in the next
section that there is actually an {\em infinite dimensional}
group of isometries. 
\par

Also interesting are those transformations that leave invariant the metric up to a Weyl rescaling. Those are
the conformal isometries which in four dimensions span the group called by Penrose and Rindler 
the Newman-Unti (NU) group (cf. \cite{Penrose}), i.e.
\bea
&& x^0\rightarrow F(x^0,z,\bar{z})\nonumber\\
&& z\rightarrow \frac{a z + b}{c z + d}
\eea
The NU group is an infinite dimensional extension of the M\"obius group. 

\section{Degenerate Horospheric coordinates}
It could appear curious that when writing the metric of the cone $N_{+}$ in terms of the degenerate horospheres
as in eq. (\ref{degenerada}) translation invariance is apparent in the coordinates $(y_1,y_2)$. Physically what happens is that
those coordinates are a sort of stereographic projection, singular when $x^0=x_3$. 
The exact relationship between cartesian and horospheric coordinates in the infinite curvature limit is:
\bea
&&x_0=\frac{1}{2 z}(y_T^2+1)\nonumber\\
&&x_3=\frac{1}{2 z}(y_T^2-1)\nonumber\\
&&x_T=\frac{y_T}{z}
\eea
where the subscript {\em transverse} refers to the $(1,2)$ labels: $y_T\equiv (y_1,y_2)$. It is worth pointing out that
the coordinate $z$ has got dimensions of energy, whereas the $y_T$ are dimensionless.
\par
Horospheric coordinates then break down when $x_0=x_3$; that is, when $z=\infty$.
\par
It is a simple matter to recover the Killings corresponding to the Lorentz subgroup:
\bea
&&J_1=-z y_1 \frac{\pd}{\pd z}-\frac{1}{2}(y_1^2 -y_2^2 -1)\frac{\pd}{\pd y_1}+y_1 y_2 \frac{\pd}{\pd y_2}\nonumber\\
&&J_2=-z y_2 \frac{\pd}{\pd z}-y_1 y_2\frac{\pd}{\pd y_1}+\frac{y_1^2 -y_2^2-1}{2}\frac{\pd}{\pd y_2}\nonumber\\
&&J_3= y_2\frac{\pd}{\pd y_1}-y_1\frac{\pd}{\pd y_2}\nonumber\\
&&K_1=y_2 z\frac{\pd}{\pd z}-y_1 y_2\frac{\pd}{\pd y_1}+\frac{y_2^2 -y_1^2 -1}{2}\frac{\pd}{\pd y_2}\nonumber\\
&&K_2=- y_1 z\frac{\pd}{\pd z}-\frac{y_2^2 + 1 - y_1^2}{2}\frac{\pd}{\pd y_1}-y_1 y_ 2\frac{\pd}{\pd y_2}\nonumber\\
&&K_3= z\frac{\pd}{\pd z}+y_1\frac{\pd}{\pd y_1}+y_2\frac{\pd}{\pd y_2}\nonumber\\
\eea

But there are more Killing vectors. First of all, the two translational ones, obvious in these coordinates:
\bea
&&P_1\equiv\frac{\pd}{\pd y_1}\nonumber\\
&&P_2\equiv\frac{\pd}{\pd y_2}\nonumber\\
\eea
and some others, such as:
\bea
&&L_1=e^{y_1}\bigg(\cos{y_2} (z\frac{\pd}{\pd z}+\frac{\pd}{\pd y_1})+\sin{y_2}\frac{\pd}{\pd y_2}\bigg)\nonumber\\
&&L_2=e^{y_1}\bigg(\sin{y_2} (z\frac{\pd}{\pd z}+\frac{\pd}{\pd y_1})-\cos{y_2}\frac{\pd}{\pd y_2}\bigg)\nonumber\\
&&J_3=e^{y_2}\bigg(\cos{y_1} (z\frac{\pd}{\pd z}+\frac{\pd}{\pd y_2})+\sin{y_1}\frac{\pd}{\pd y_1}\bigg)\nonumber\\
&&L_4=e^{y_2}\bigg(\sin{y_1} (z\frac{\pd}{\pd z}+\frac{\pd}{\pd y_2})-\cos{y_1}\frac{\pd}{\pd y_1}\bigg)\nonumber\\
&&L_5=e^{-y_1}\bigg(\cos{y_2} (z\frac{\pd}{\pd z}-\frac{\pd}{\pd y_1})+\sin{y_2}\frac{\pd}{\pd y_2}\bigg)\nonumber\\
&&L_6=e^{-y_1}\bigg(\sin{y_2} (z\frac{\pd}{\pd z}-\frac{\pd}{\pd y_1})-\cos{y_2}\frac{\pd}{\pd y_2}\bigg)\nonumber\\
&&L_7=e^{-y_2}\bigg(\cos{y_1} (z\frac{\pd}{\pd z}-\frac{\pd}{\pd y_2})+\sin{y_1}\frac{\pd}{\pd y_1}\bigg)\nonumber\\
&&L_8=e^{-y_2}\bigg(\sin{y_1} (z\frac{\pd}{\pd z}-\frac{\pd}{\pd y_2})-\cos{y_1}\frac{\pd}{\pd y_1}\bigg)
\eea
More Killings are gotten through commutation; the boost $K_3$, in particular, raises powers of the coordinates when acting on the
$L$'s:
\bea
&&[K_3,L_1]= e^{y_1}\bigg((y_1\cd-y_2\sd)z\frac{\pd}{\pd z}+((y_1-1)\cd-y_2\sd)\frac{\pd}{\pd y_1}\nonumber\\
&&+((y_1-1)\sd+y_2\cd)\frac{\pd}{\pd y_2}\bigg)\equiv Q_1\nonumber\\
&&[K_3,L_2]= e^{y_1}\bigg((y_1\sd-y_2\cd)z\frac{\pd}{\pd z}+(-(y_1-1)\cd+y_2\sd)\frac{\pd}{\pd y_2}\nonumber\\
&&+((y_1-1)\sd+y_2\cd)\frac{\pd}{\pd y_1}\bigg)\equiv Q_2\nonumber\\
\eea
Clearly the process never ends. Commuting again with $K_3$ produces terms in $y_1^2 e^{y_1}$ which are not found amongst
the existing generators. The isometry group is then infinite dimensional.
\par

It is actually possible to give the general solution of the Killing equation in closed form using horospheric 
coordinates. Given an arbitrary analytic
function of the complex variable $y_1+i y_2$, for example $f(y_1+i y_2)$, it is given by:
\be
k\equiv (\frac{\pd^2}{\pd y_1^2}Re\, f)z\frac{\pd}{\pd z}+(\frac{\pd}{\pd y_1}Re\, f) \frac{\pd}{\pd y_1}
-(\frac{\pd}{\pd y_2}Re\, f) \frac{\pd}{\pd y_2}
\ee

\par
It is now clear that the isometry group of the four-dimensional light cone $N_{+}$ is an infinite dimensional group, 
which includes the Lorentz group as a subgroup.
\par

We find this to be a remarkable situation.

Even more remarkable is the fact that in higher dimension, when the total space gets dimension
$d$, say, so that the light cone has dimension $d-1$, and in horospheric coordinates is
 characterized by $z$ and $\vec{y}\in\mathbb{R}^{d-2}$, the Killing equations are
equivalent to
\be\label{conf}
\pd_i k_j+\pd_j k_i = 2\d_{ij}k(\vec{y})
\ee 
for the total vector
\be
k=z k(\vec{y}) \pd_z + \sum_{i=1}^{d-2} k^i\pd_i
\ee
But the equations (\ref{conf}) are precisely the equations for the conformal Kiling
vectors of flat $(d-2)$-dimensional space, known to generate the euclidean 
conformal group, $SO(1,d-1)$, isomorphic to the $d$-dimensional Lorentz group. 
To be specific (\cite{Erdmenger}),
\be
k(\vec{y})\equiv \lambda-2 \vec{b}.\vec{y}
\ee
and the components on the $y$-directions read:
\be
k_i=a_i+\omega_{ij} y^j+\lambda y_i + b_i y^2-2 \vec{b}.\vec{y} y_i
\ee

representing translations ($a$), rotations ($\omega_{(ij)}=0$), scale transformations, 
$(\l)$, and 
special conformal transformations ($b)$.

\par
To summarize, the isometry group of the light cone at the origin, $N_{+}(0)$, is generically
the spacetime Lorentz group {\em except} in the four dimensional case, in which it expands
to the infinite group  we derived above.

\section{Conclusion: sigma models on singular manifolds.}
We can expect this approximation to work for lengh scales much larger than 
the one defined by the curvature inverse, i.e. it is a low energy approximation, valid for $E<< l^{-1}$.

The (singular) propagator boundary-boundary 
we get in this way (when $l\equiv \e\rightarrow 0$) is:
\be
\Delta_{b-b}\equiv\frac{\e^{n+1}\Gamma(n-1)}{\pi^{(n-1)/2}\Gamma(\frac{n-1}{2})}\,\frac{z^{n-1}}{|\vec{y}-
\vec{y}^{\prime}|^{n-2}}
\ee

It is remarkable that this propagator is explicitly translationally invariant, in spite 
of the fact that the light cone is topologically a sphere with time-dependent radius.
\par
It is quite difficult however to make use of this fact in order to progress along these lines 
in a sigma
 model approach.
For example, the usual representation of $AdS_3$ as a Wess-Zumino-Witten (WZW)
model leads to the lagrangian:
\be
L=2k\bigg(\frac{1}{u^2}\pd u \bar{\pd}u +u^2\pd \bar{\g} \bar{\pd}\g\bigg)
\ee
(where $(u,\g,\bar{\g})$ are coordinates descibed in detail in \cite{Giveon}).
The parameter 
\be
k= l^2
\ee
so that in the degenerate limit $k=0$. But this is bad, because the central charge
of the underlying CFT is
\be
c=\frac{3 k}{k-2}
\ee
so that usual considerations are restricted to $k>2$. More work on these issues
 can be, however, rewarding.
\section*{Acknowledgments}
I am grateful to Edgar Gunzig and Enric Verdaguer for the invitation to the marvellous site of Peyresq,
 and to Jaume Garriga and Enric Verdaguer for useful discussions.

This work ~~has been partially supported by the
European Commission (HPRN-CT-200-00148) and FPA2003-04597 (DGI del MCyT, Spain).


\end{document}